\documentclass{iau}
\usepackage{graphicx}

\title[IAUS 289.~~The young open cluster Westerlund 2] 
{The distance to the young open cluster Westerlund 2}

\author[G. Carraro, D. Turner, D. Majaess, G. Baume]   
{Giovanni Carraro$^1$,
David Turner$^2$, Daniel Majaess$^2$, Gustavo Baume$^3$}

\affiliation{$^1$ESO, Alonso de Cordova 3107, 19001, Santiago de
  Chile, Chile \\ email: {\tt gcarraro@eso.org} \\[\affilskip]
$^2$ Department of Astronomy and Physics, Saint Mary’s University,
Halifax, NS B3H 3C3, Canada\\email: {\tt turner@ap.smu.ca}\\[\affilskip]
$^3$ Facultad de Ciencias Astron\'omicas y Geof\'isicas (UNLP), Instituto
de Astrof\'isica de La Plata (CONICETUNLP), Paseo del Bosque s/n, La
Plata,  Argentina\\email: {\tt gbaume@fcaglp.unlp.edu.ar}}

\pubyear{2013}
\volume{289}  
\pagerange{119--126}
\jname{Advancing the physics of cosmic distances}
\editors{R. de Grijs \& G. Bono , eds.}
\begin{document}

\maketitle

\begin{abstract}
Evidence is presented indicating that the young star cluster Westerlund~2 lies $d_{\odot}\sim3.0$ kpc in the direction of Carina. 
The distance is tied partly to new $UBVRI_c$ photometry and revised spectral classifications for cluster stars, which imply that dust in the direction of Carina is characterized by an anomalous extinction law ($R_V\sim3.8$).  That result was determined from a multi-faceted approach relying on the variable-extinction and color excess methods. 
\end{abstract}


\section{Introduction}
Westerlund 2 is a compact young open cluster located in the direction of the Carina spiral arm ($\ell,b=284.4^{\circ},-0.34 ^{\circ}$).  The distance for Westerlund 2 has been the subject of lively debate (see \cite[Rauw et
al. 2007]{ra07} and \cite[Ascenso et al. 2007]{As07}), with estimates ranging from 2.5 to more than 8 kpc from the Sun.  In this analysis, we employ new $UBVRI_c$  photometry to infer the cluster distance.  The revised results have pertinent ramifications for the determination of the cluster mass, and the reputed cluster membership of WR20aa and WR20c (\cite[Roman-Lopes et al. 2011]{rl11}).\\
\noindent
The photometric data were acquired via the SWOPE telescope, which is stationed on Las Campanas.  Details of the photometric reduction will be presented elsewhere (Carraro et al. 2012, in preparation).  Instead of assuming that the canonical extinction law ($R_V$, the ratio of the total to selective absorption $A_V/E(B-V)$) characterizes dust associated with Westerlund 2, as routinely done in previous studies, the parameter is derived using a combination of optical and infrared photometry. The distance to Westerlund 2 is acutely sensitive to the value of $R_V$ owing to the large reddening found for cluster stars ($\mu_0=V-M_V-R_v \times E(B-V)$).  The variable-extinction and color excess methods were both employed to infer $R_V$, and 
point to an anomalous value of $R_V\sim3.8$.\\
\noindent \\
In this contribution we briefly summarize our results.

\section{The Distance to Westerlund~2}
Any determination of the distance to Westerlund~2 is intimately tied to how interstellar reddening and extinction affect cluster
stars. That is particularly important for Westerlund~2 because of its 
location in the vicinity of the Great Carina Nebula, where anomalous
extinction dominates (\cite[Carraro et al. 2004]{Ca04},  \cite[Turner (2012)]{tu12}). 
The CCD photometry presented here was therefore used to test the
reddening properties of the dust associated with 
Westerlund~2.  Spectra for cluster O-type stars
published by \cite[Rauw et al. 2007]{ra07} were reanalyzed to confirm their 
temperature and luminosity subtypes on the Walborn system
(\cite[Walborn 1971a,1971b,1972,1990]{wa71a,wa71b,wa72,wp90}).
Colour excesses, $E_{B-V}$ and $E_{U-B}$, were then calculated for the stars with reference
to an unpublished set of intrinsic colours for early-
type stars established by DGT (\cite[Turner et al. 2012]{te12}).  The latter was compiled in part by melding
published values cited by \cite[Johnson (1966)]{jo66} and \cite[FitzGerald (1970)]{fi70}. 
The results are presented in Table \ref{tab1}, where the star
numbering is from \cite[Moffat et al. 1991]{mo91}.

\setcounter{table}{0}
\begin{table}
\caption[]{Data for Bright Members of Westerlund 2.}
\label{tab1}
\begin{center}
\begin{tabular}{@{\extracolsep{0pt}}cccclccc}
\hline
MSP &{\it V} &{\it B--V} &{\it U--B} &Sp.T. &$E_{B-V}$ &$E_{U-B}$ &{\it V--M}$_V$ \\
\hline
18 &12.79 &1.32 &--0.17 &O3~III((f)) &1.65 &1.03 &18.69 \\
151 &14.24 &1.50 &--0.11 &O5~V((f)) &1.83 &1.07 &19.34 \\
157 &14.08 &1.45 &--0.03 &O6~V((f)) &1.78 &1.15 &19.08 \\
167 &14.08 &1.64 &--0.09 &O4~III(f) &1.97 &1.10 &19.88 \\
171 &14.30 &1.65 &+0.03 &O3~V((f)) &1.98 &1.23 &19.70 \\
175 &13.86 &1.33 &--0.03 &O4~V((f)) &1.66 &1.16 &19.06 \\
182 &14.28 &1.46 &--0.07 &O3~V((f)) &1.79 &1.13 &19.68 \\
183 &13.48 &1.70 &+0.05 &O2~V(f*) &2.03 &1.25 &20.08$^{\rm a}$ \\
188 &13.27 &1.49 &--0.06 &O4~III &1.82 &1.13 &19.68 \\
199 &14.23 &1.55 &+0.02 &O3~V((f)) &1.88 &1.22 &19.63 \\
203 &13.20 &1.52 &--0.06 &O3~V((f))n &1.85 &1.14 &19.10$^{\rm a}$ \\
263 &14.79 &1.85 &+0.14 &O3~V((f)) &2.18 &1.34 &20.69$^{\rm a}$ \\
\hline
\end{tabular} \\
Notes: $^{\rm a}$ Luminosity for class IV adopted.
\end{center}
\end{table}

\begin{figure}[!t]
\includegraphics[width=7.3cm]{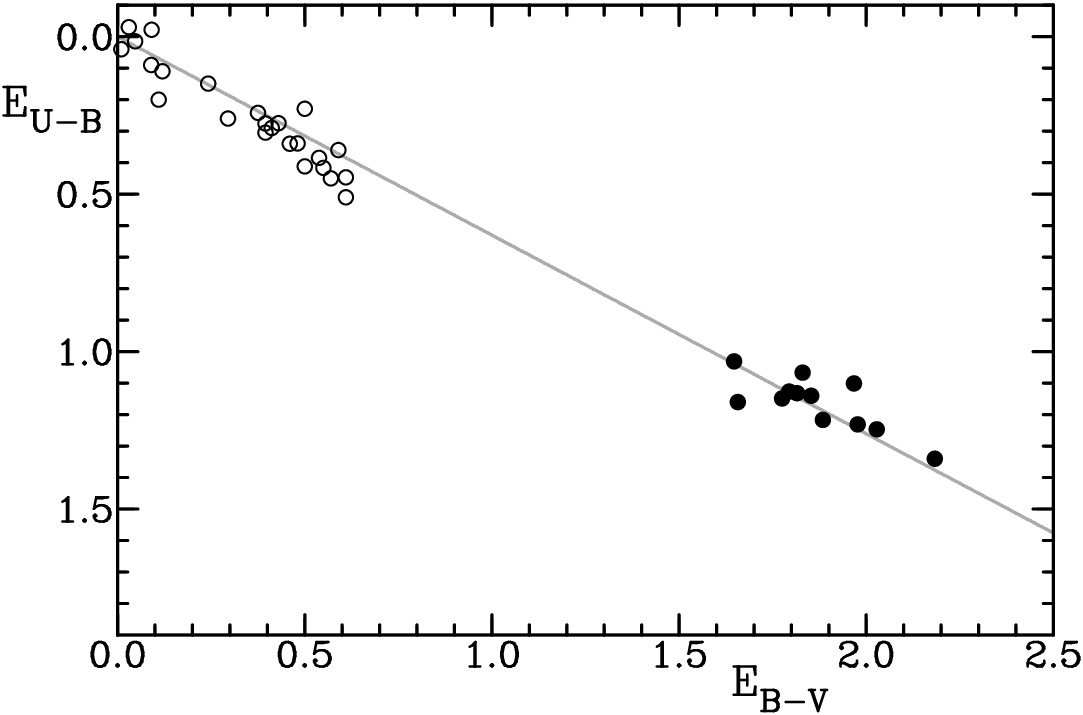}
\includegraphics[width=5.5cm]{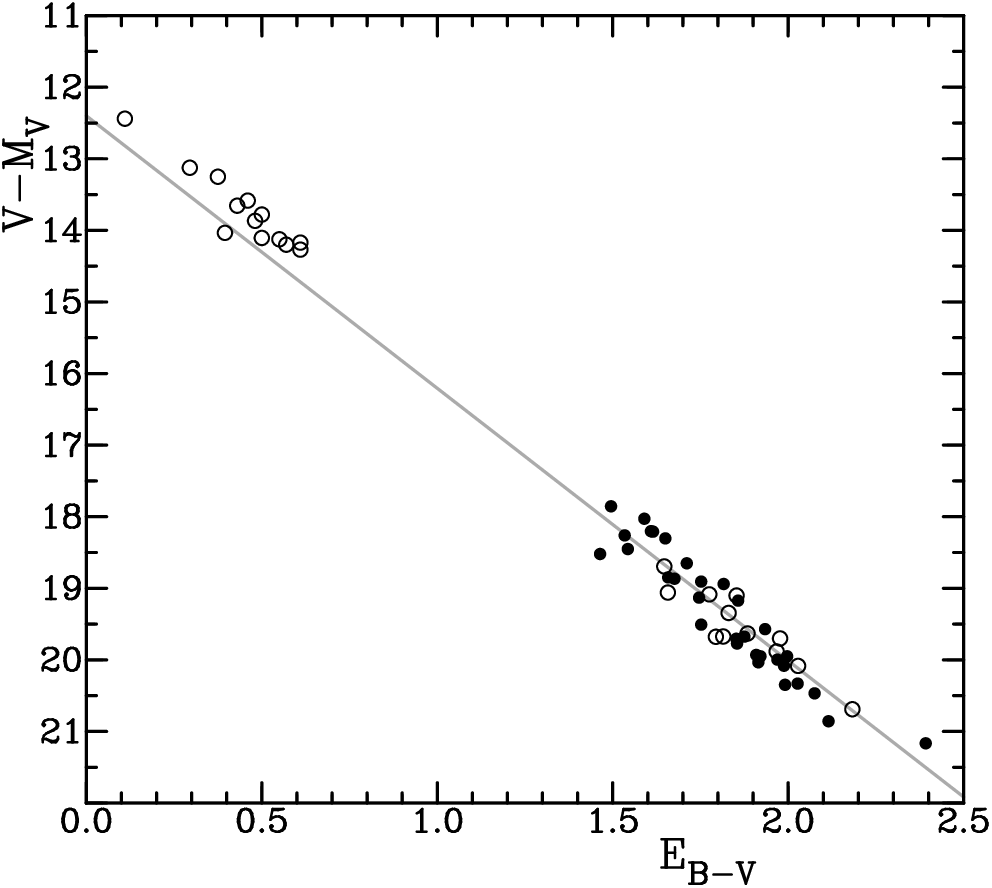}
\caption{ {\bf Left panel}: a plot of {\it UBV} colour excesses for
  spectroscopically observed stars in Westerlund~2 (filled circles)
  lying $r<1^{\circ}$ from the cluster (open circles). The reddening relation
  plotted in gray is described by $E_{U-B}/E_{B-V}=0.63 +
  0.02\;E_{B-V}$. {\bf Right panel}: a variable-extinction diagram for members of Westerlund~2
  (right) inferred from ZAMS fitting (filled circles) and spectroscopic
  distance moduli (open circles). Stars around IC~2581 constitute the
  group exhibiting smaller reddenings. The gray line represents the best
  fitting parameters: $V_0-M_V=12.40$ and $R_V=A_V/E_{B-V}=3.80$ }
\label{fig1}
\end{figure}

Colour excesses were also calculated for other stars lying within
$1^{\circ}$ of Westerlund~2 using data from the literature, with the
results depicted in Fig.\,\ref{fig1} (left panel). The data for Westerlund 2 exhibit some scatter
owing in part to {\it U--B} uncertainties, and for surrounding stars partly because of variations in the
extinction properties of dust affecting individual stars (\cite[Turner
(2012)]{tu12}). 
Overall, however, there is excellent agreement with a reddening
relation described by $\frac{E_{U-B}}{E_{B-V}}=0.63 + 0.02\;E_{B-V}$, where
the reddening slope matches 
that found by \cite[Turner (2012)]{tu12} for dust affecting the heavily-reddened
group of stars associated with WR38.  WR38 is also in the Great Carina
Nebula complex.  The curvature term is adopted from \cite[Turner (1989)]{tu89}. Only 3 of the 12
spectroscopically-observed stars deviate from the adopted relation. The stars are faint and the
photometry may be affected by crowding.

CCD {\it UBV} data for other cluster stars were used in a
variable-extinction analysis using zero-age
main-sequence (ZAMS) fitting 
techniques (\cite[Turner (1976a,b)]{tu76a,tu76b}).  The ZAMS for O-type stars
is that of \cite[Turner (1976a)]{tu76a}, updated to include the hottest O-type stars. 
Spectroscopically-observed stars anchored the analysis through their
apparent distance moduli (Table~\ref{tab1}), although adjustments were 
necessary for the luminosity classifications of three stars so
they did not deviate from other cluster members. The luminosity
calibration is tied to the same ZAMS relation (\cite[Turner (1980)]{tu80}). The analysis was restricted to
ZAMS-fitted stars, producing the results depicted in Fig.\,\ref{fig1} (right panel).
The best-fitting values obtained from least squares and non-parametric
techniques are $R_V=A_V/E_{B-V}=3.80\pm0.20$ and $V_0-M_V=12.40\pm0.37$, and
correspond to a distance of $3.02\pm0.52$ kpc. The large
uncertainty in the distance arises from the large reddening, and the
extrapolation 
from a reddening of $E_{B-V}\simeq1.75$. Note that a large value of
$R_V$ is consistent with the shallow reddening slope, as found for
adjacent 
regions of Carina (\cite[Turner  (2012), Turner et al. 2009,
Turner et al. 2005]{te05,te09,tu12}). A similar analysis for stars
in Fig.\,\ref{fig1} associated with IC~2581 yields a value of 
$R_V=3.77\pm0.19$ and $V_0-M_V=12.01\pm0.09$, corresponding to a
distance of $2.52\pm0.11$ kpc. 
Many of the other clusters in the region of the Great Carina Nebula
converge towards distances of $\sim2.1$ kpc (\cite[Turner (2012)]{tu12}). 
Westerlund~2 appears to lie on the far side of the group.

\begin{figure}[!t]
\includegraphics[width=7.5cm]{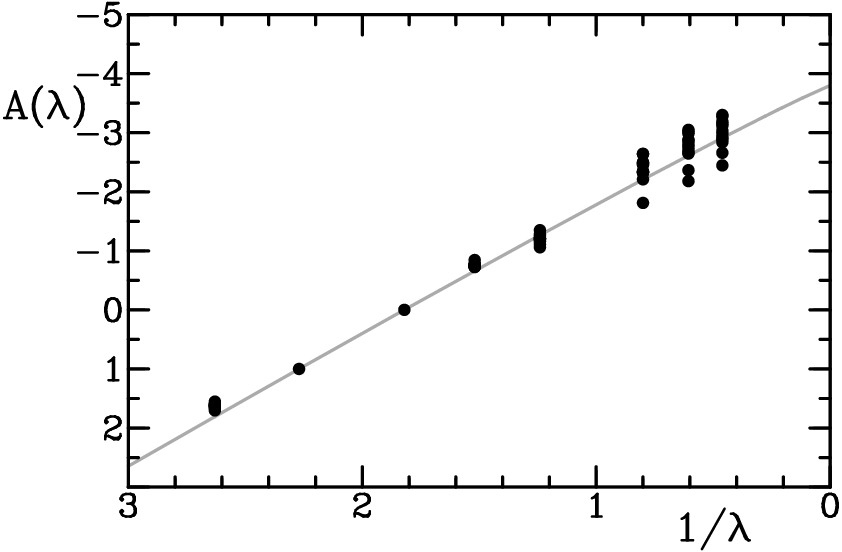}
\includegraphics[width=4.5cm]{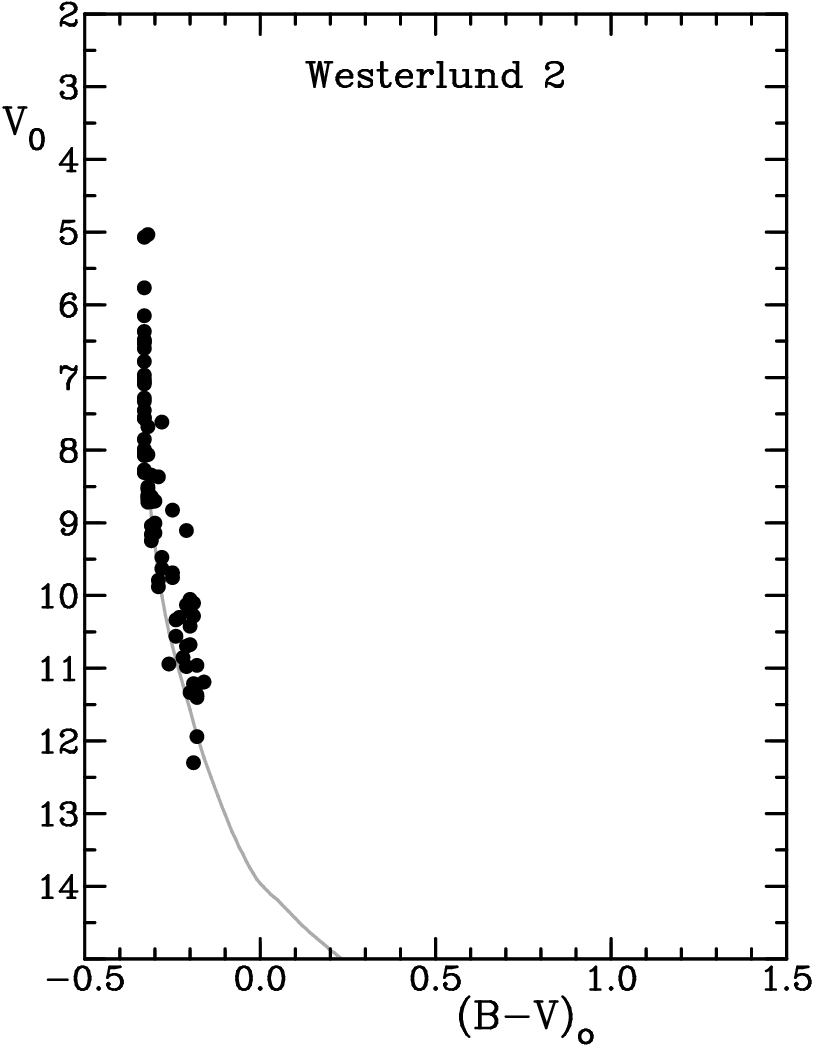}
\caption{{\bf Left panel}: colour-excess method applied to the $UBVRI_cJHK_s$
  photometry for Westerlund~2 stars. The fitted relation implies $R_V=A_V/E_{B-V}=3.8$.
{\bf Right panel}: Unreddened colour-magnitude diagram for Westerlund~2 for the parameters cited in the text. The gray line represents the ZAMS for $V_0-M_V=12.40\pm0.37$. $M_{\odot}$}
\label{fig2}
\end{figure}

The large value of $R_V$ obtained from the variable-extinction analysis
is confirmed by the colour difference 
method using $UBVRI_c$ data from this study, in conjunction with {\it
  JHK}$_s$ data from 2MASS (\cite[Cutri et al. 2003]{cu03}).  Intrinsic {\it JHK}$_s$ 
colours for hot O-stars were adopted from \cite[Turner (2011)]{tu11}. The results
depicted in Fig.\,\ref{fig2} (left panel) yield an average value of $R_V=3.80\pm0.09$
for the 12 spectroscopically-observed stars, using the relationship of
\cite[Fitzpatrick \& Massa (2007)]{fm07}. The function fitted to the data in
Fig.\,\ref{fig2} (left panel) is adapted from \cite[Zagury \& Turner (2012)]{zt12} for that value of $R_V$.

The unreddened colour-magnitude diagram for Westerlund~2
(Fig.\,\ref{fig2}, right panel) describes a rather young cluster. 
There are essentially no hot O-type stars evolved significantly from
the ZAMS, indicating that the cluster contains no O-type supergiants. 
Standard stellar evolutionary models (\cite[Meynet et
al. 1993]{me93}) imply an age of no more than $\sim2\times10^6$ yr for cluster stars.

\section{Conclusions}
In this contribution we determined that the extinction law toward
Westerlund 2 is anomalous ($R_V \sim 3.8$).
We therefore propose that Westerlund 2 lies $d\sim3.0$ kpc from
the Sun in the direction of Carina.  The results confirm the short-distance to Westerlund 2 advocated by \cite[Ascenso et al. 2007]{As07} on the basis of near-infrared $JHK_s$ data, and rules out larger distances.  The cluster distance inferred from a PMS/MS fit to X-ray (Chandra) cleaned $VI_cJHK_s$ color-magnitude diagrams corroborates that conclusion (Carraro et al. 2012, in preparation).

\end{document}